\documentclass[useAMS,usenatbib,fleqn]{mnras}

\usepackage{graphicx,color,amsmath,amssymb,amsfonts}

\newcommand{\be}{\begin{equation}}
\newcommand{\ee}{\end{equation}}
\newcommand{\bea}{\begin{eqnarray}}
\newcommand{\eea}{\end{eqnarray}}

\newcommand{\dco}{\delta_{\rm cs}}
\newcommand{\de}{\mathrm d}

\newcommand{\om}{\Omega_m}
\newcommand{\ob}{\Omega_b}
\newcommand{\odm}{\Omega_c}

\newcommand{\ho}{H_0}

\newcommand{\lcdm}{$\Lambda$CDM\,}

\newcommand{\fnl}{{f_\mathrm{NL}}}
\newcommand{\fng}{{f_\mathrm{NL}^{\rm GR}}}
\newcommand{\fnc}{{f_\mathrm{NL}^{\rm CMB}}}
\newcommand{\fns}{{f_\mathrm{NL}^{\rm LSS}}}
\newcommand{\mb}{\mathcal{Q}}
\newcommand{\lm}{{\ell_{\rm min}}}
\newcommand{\lM}{{\ell_{\rm max}}}
\newcommand{\fsky}{{f_{\rm sky}}}

\title[Probing primordial non-Gaussianity with SKA galaxy redshift surveys]{Probing primordial non-Gaussianity with SKA galaxy redshift surveys: a fully relativistic analysis}
\author[S. Camera, M.G. Santos \& R. Maartens]{Stefano Camera,$^{1,2,3}$\thanks{E-mail: stefano.camera@manchester.ac.uk.} M\'ario G. Santos$^{3,4,2}$ \& Roy Maartens$^{3,5}$\\
$^1$Jodrell Bank Centre for Astrophysics, School of Physics and Astronomy, The University of Manchester, Manchester M13 9PL, UK\\
$^2$CENTRA, Instituto Superior T\'ecnico, Universidade de Lisboa, 1049-001 Lisboa, Portugal\\
$^3$Physics Department, University of the Western Cape, Cape Town 7535, South Africa\\
$^4$SKA SA, The Park, Park Road, Cape Town 7405, South Africa\\
$^5$Institute of Cosmology \& Gravitation, University of Portsmouth,  Portsmouth PO1 3FX, UK}
\date{}

\begin{document}

\date{Accepted 0000 --- 00. Received 0000 --- 00; in original form 0000 --- 00}

\pagerange{\pageref{firstpage}--\pageref{lastpage}} \pubyear{2014}

\maketitle

\label{firstpage}
\begin{abstract}
The Square Kilometre Array (SKA) will produce spectroscopic surveys of tens to hundreds of millions of neutral hydrogen (HI) galaxies, eventually covering 30,000 deg$^2$ and reaching out to redshift $z\gtrsim2$. The huge volumes probed by the SKA will allow for some of the best constraints on primordial non-Gaussianity, based on measurements of the large-scale power spectrum. We investigate various observational set-ups for HI galaxy redshift surveys, compatible with the SKA Phase 1 and Phase 2 (full SKA) configurations. We use the corresponding number counts and bias for each survey from realistic simulations and derive the magnification bias and the evolution of source counts directly from these. For the first time, we produce forecasts that fully include the general relativistic effects on the galaxy number counts. These corrections to the standard analysis become important on very large scales, where the signal of primordial non-Gaussianity grows strongest. Our results show that, for the full survey, the 
non-Gaussianity parameter $\fnl$ can be constrained down to $\sigma(\fnl)=1.54$. This improves the current limit set by the \textit{Planck} satellite by a factor of five, using a completely different approach.
\end{abstract}

\begin{keywords}
cosmology: large-scale structure of the universe---early Universe---cosmological parameters---observations---radio lines: galaxies---relativistic processes.
\end{keywords}

\section{Introduction}
The large-scale distribution of matter contains information not only about the current state of the Universe---such as the dark energy---but also about the primordial state of the Universe, at times well before the onset of structure formation. This remarkable fossil record is imprinted in the correlations of galaxies by the primordial fluctuations which evolve to become the seeds of structure, and it contains vital clues as to the inflationary mechanism (or an alternative mechanism) for generating these fluctuations. 

One of the most important questions is whether or not the primordial fluctuations are Gaussian. Primordial non-Gaussianity (PNG) imprints a characteristic feature, via the bias $b$, in the galaxy power spectrum $P_g=b^2P_m$. (It also imprints features in the bispectrum and higher moments, which we do not consider here.) This feature causes a growth of power $\propto k^{-2}$ on large scales. The excess power is `frozen' on super-Hubble scales during the evolution of the galaxy overdensity---it is not affected by non-linearity on small scales. By contrast, PNG is continually generated on small scales by structure formation so that the record of PNG can be washed out on these scales.

The best current constraints on PNG are from cosmic microwave background (CMB) measurements by the \textit{Planck} satellite \citep{Ade:2013ydc}. In particular, for the local form of PNG (which has the strongest impact on galaxy bias) we have
 \be\label{fp}
 \fnc = 2.7\pm 5.8 ~~(1\sigma).
 \ee
This result already rules out inflationary models with large PNG. In order to discriminate amongst the remaining models we need to bring down the errors. Ultimately, we need to probe the simplest single-field models that generate negligible PNG. [Note that there is a subtle point here, which is discussed in the next section; see in particular Eq.~\eqref{fidg}.]
 
Galaxy surveys currently lag well behind \textit{Planck}, but they can provide a valuable consistency check by probing PNG at low redshifts  \citep[see e.g.][for the current level of constraints on $\fnl$]{Ross:2012sx,Giannantonio:2013uqa}. Furthermore, future surveys covering a large fraction of the sky, such as those planned by the Square Kilometre Array (SKA),\footnote{\texttt{http://www.skatelescope.org/}} will be able to probe PNG over a significant range of redshifts, giving a much higher number of modes than the CMB, which is confined effectively to a single redshift. If systematics and complications in the  modelling of bias and haloes can be controlled, then the future of PNG constraints lies with huge-volume galaxy surveys.

Here, we investigate the constraining power of future neutral hydrogen (HI) galaxy redshift surveys with the SKA, which are envisaged to cover 3/4 of the sky and reach a depth of $z\sim2$ in the full SKA. We use the Fisher forecast method, where we include the relevant cosmological parameter $\sigma_8$. We carefully include the effects of magnification bias and evolution of source counts, based on simulations of HI galaxy number densities. In addition, we include all relativistic effects that enter into the theoretical modelling of the {\it observed} power spectra. This removes an important theoretical systematic which is present if one neglects the relativistic effects.

We assume a `concordance' (flat \lcdm) model for the background.

\section{PNG with relativistic effects}\label{sec:png-gr}
If the distribution of primordial curvature perturbations is not Gaussian, it cannot be fully described by a power spectrum $P_\Phi(k)$, where $\Phi$ is Bardeen's gauge invariant potential. We need higher-order moments, starting with the bispectrum $B_\Phi(\mathbf k_1,\mathbf k_2,\mathbf k_3)$, where $\mathbf k_1+\mathbf k_2+\mathbf k_3=0$. Different models of inflation give rise to different shapes of the bispectrum. For the effects of PNG on galaxy bias, the `local' (or `squeezed') configuration is dominant on large scales \citep{Taruya:2008pg,Fedeli:2009fj}. In this case, the bispectrum is maximised for triangles with one of the three momenta much smaller than the other two. Simple single-field inflation models generate local PNG. There are inflationary models that generate different shapes for the primordial bispectrum, e.g. `equilateral' \citep{Crociani:2008dt,Fedeli:2009fj} and `folded' \citep{Holman:2007na,Meerburg:2009ys,Verde:2009hy}. However, the local shape gives the largest effects---particularly 
on bias---and  we confine our analysis to local PNG, i.e. $\fnl=f_{\rm NL}^{\rm local}$ from now on.

A convenient way to parametrise the deviation from Gaussianity is  \citep{Salopek:1990jq,Gangui:1993tt,Verde:1999ij,Komatsu:2001rj}
\begin{equation}\label{eqn:ng}
\Phi=\varphi-\fnl\left(\varphi^2-\langle\varphi^2\rangle\right),
\end{equation}
where $\varphi$ is a linear Gaussian potential. PNG modulates the formation of haloes in the cold dark matter, inducing a scale (and redshift) dependence in the halo bias \citep{Dalal:2007cu,Matarrese:2008nc}. The modification  to the Gaussian bias on large scales is given by
\begin{align}
b(z) & \to b(z)+\Delta b(z,k),\\
\Delta b(z,k)&=[b(z)-1]\frac{3\om\ho^2q\delta_{\rm cr}}{k^2T(k)D(z)}\fnl, \label{eq:bias-NG}
\end{align}
where the factor $q=\mathcal O(1)$ is included to obtain agreement with $N$-body simulations \citep{Carbone:2008iz,Grossi:2009an,Wagner:2011wx}. Given the uncertainty in modelling $q$ and in extracting it reliably from $N$-body simulations, we follow \citet[][]{Giannantonio:2011ya}, and most works that constrain $\fnl$ via the bias, in taking $q=1$. This means that we in fact constrain an effective $f_{\rm NL}^{\rm eff}=q\times\fnl$ until such time as the theoretical uncertainty in $q$ is resolved. In Eq. (4),  $\om=\ob+\odm$ is the total matter fraction at $z=0$, $\ho$ is the Hubble constant, $\delta_{\rm cr}\simeq1.69$ is the critical matter density contrast for spherical collapse, $T(k)$ is the matter transfer function ($T\to1$ as $k\to0$) and $D(z)$ is the linear growth function of density perturbations (normalised to unity today). The most important feature  in Eq.~\eqref{eq:bias-NG} is the $k^{-2}$ term, which comes from relating $\Phi$ to the overdensity $\delta$ via the Poisson equation. This term means that the PNG signal in the bias grows stronger as $k\to0$. Note that $\Delta b$ is strongly suppressed on scales smaller than the equality scale, so that even if we include such scales in computations, there is no false gain in constraining power over $\fnl$ \citep[see also][]{Giannantonio:2011ya}.

In Eq.~\eqref{eq:bias-NG} we have used the large-scale structure (LSS) convention  for defining $\fnl$, as opposed to the CMB convention  \citep{Afshordi:2008ru,2010MNRAS.402..191P,Carbone:2008iz,Grossi:2009an}. The difference arises since $\Phi$ is evaluated at decoupling for $\fnc$ and at low $z$ for $\fns$. The relation is
\be
\fns=\frac{g(z_{\rm dec})}{g(0)}\,\fnc\simeq\frac{1}{5}\big(3+2\Omega_m^{-0.45}\big) \fnc \simeq 1.3 \fnc,
 \label{ctl}
\ee
where $g(z)=(1+z)D(z)$. This means that when we constrain PNG using Eq.~\eqref{eq:bias-NG}, the \textit{Planck} constraint of Eq.~\eqref{fp}, given in the CMB convention, translates in the LSS convention to
\be\label{fpg}
 \fnl = 3.5\pm 7.5 ~\Rightarrow~ \sigma(\fnl)_{\rm Planck}=7.5\,.
 \ee
Here and in the rest of this paper, $\fnl$ is in the LSS convention, i.e. $\fnl=\fns$. Thus the target for galaxy surveys to beat is  $\sigma(\fnl)=7.5$.

The standard approach to constraining PNG via the galaxy bias proceeds from Eq.~\eqref{eq:bias-NG}, and uses  power spectrum measurements that include redshift space distortions (RSD) via the Kaiser term,
\be\label{dk}
\delta_g^z = \delta_g -\frac{(1+z)}{{H}} (n^i\partial_i)^2 V ~\mbox{where}~v_i=\partial_i V.
\ee
Here, $n^i$ is the direction of the galaxy and $v^i$ is its peculiar velocity. In addition, the fiducial value for forecasting constraints on PNG is taken as
\be
f_{\rm NL}^{\rm fid}=0. \label{fid}
\ee
This approach is reasonable for large $\fnl$, which overwhelms any relativistic corrections. But \textit{Planck} has shown $\fnl$ is not large---and so we need to be more careful and include all relativistic effects.

These effects are of two kinds:\\ (1) a non-linear primordial correction; \\ (2) linear projection effects from observing in redshift space on the past light-cone.

\subsection{Primordial relativistic correction}
In the Newtonian approximation, an exactly Gaussian distribution of the primordial curvature perturbation translates into an exactly Gaussian distribution of density perturbations via the Newtonian Poisson equation, $\nabla^2\Phi=4\pi G a^2\rho\delta$. In general relativity however, the Newtonian Poisson equation is not correct at second order. There is a relativistic non-linear correction to the Poisson constraint that links curvature to density perturbations. As a result, an exactly Gaussian distribution of primordial curvature perturbations does not lead to a Gaussian distribution of density perturbation, even on super-Hubble scales.

This  non-linearity was first derived in \cite{Bartolo:2005xa}, and the implications for PNG were then developed in \cite{Verde:2009hy} \citep[see also][]{Hidalgo:2013mba,Bruni:2013qta,Villa:2014foa}. The effective local PNG parameter that describes this primordial general relativistic correction on large scales is
\be \label{fgr}
\fng = -\frac{5}{3} ~~\mbox{(CMB convention)}.
\ee
This correction is derived in the CMB convention because it is based on the primordial $\Phi$. It does not affect CMB measurements of PNG, because these are independent of the Poisson constraint. But for large-scale structure, it must be added to the local PNG parameter. Translating Eq.~\eqref{fgr} to the LSS convention via Eq.~\eqref{ctl}, we have
\be
\fnl \to \fnl+\fng \simeq \fnl -2.17\,. \label{fcor}
\ee

Equation~\eqref{fcor} has interesting implications. Firstly, it means that the standard fiducial value of Eq.~\eqref{fid} should be corrected to
\be
f_{\rm NL}^{\rm fid}\simeq -2.17\,. \label{fidg}
\ee
If the primordial curvature perturbation is Gaussian, which is effectively the case for slow-roll single-field inflation, then this is the local PNG that we expect to measure in galaxy surveys. {\it For the concordance model with slow-roll single-field inflation, galaxy surveys should measure $\fnl\simeq-2.17$.} This subtle point \citep[see][]{Verde:2009hy} is missed in most previous work on PNG from single-field inflation \citep[e.g.][]{Pajer:2013ana}. 

Finally, Eq.~\eqref{fidg} implies that the necessary accuracy for testing single-field inflation through PNG in galaxy surveys is $\sigma(\fnl)\lesssim2$.

\subsection{Relativistic projection effects}\label{ssec:gr}
The Kaiser RSD term in Eq.~\eqref{dk} is the dominant term on sub-Hubble scales of a more complicated set of relativistic terms that arise from radial and transverse perturbations along the light-ray of observation. (Note that these terms are at linear order, unlike the non-linear correction in the previous subsection.) 

The first such term is the contribution of the lensing convergence $\kappa$ to the observed number density contrast, where 
\be
\kappa =\int_0^\chi d\tilde\chi\,(\chi-\tilde\chi)\frac{\tilde\chi}{\chi}\nabla_\perp^2 \Phi. \\
\ee
Here $\chi$ is the radial comoving distance, related to the Hubble rate $H(z)$ by $\de\chi=\de z/H(z)$, and $\nabla_\perp^2$ is the Laplacian on the transverse screen space. Gravitational lensing affects the observed number density in two competing ways---enhancing it by bringing faint galaxies into the observed patch, and decreasing it by broadening the area of the patch. The competition is controlled by the magnification bias $\mb$, i.e.\ the slope of the (unlensed) galaxy luminosity function at the survey flux limit. It is given by
\begin{equation}
\mathcal Q=\left.-\frac{\partial\ln N_g}{\partial\ln\mathcal F}\right|_{\mathcal F=\mathcal F_\ast},\label{mbq}
\end{equation}
where $N_g(z,\mathcal F>\mathcal F_{\ast})$ is the background galaxy number density at redshift $z$ and with flux $\mathcal F$ above the flux detection threshold $\mathcal F_{\ast}$. Then the contribution of lensing to Eq.~\eqref{dk} is $2(\mb-1)\kappa$. 

The remaining relativistic contributions are both local and integrated terms
\citep{Yoo:2009au,Yoo:2010ni,Bonvin:2011bg,Challinor:2011bk,Jeong:2011as,Bertacca:2012tp}, namely
\be
\delta^{\rm obs}_g =  (b+\Delta b)\dco
  -\frac{(1+z)}{{H}} (n^i\partial_i)^2 V 
-2\left(1-\mathcal Q\right) \kappa + \delta_{\rm loc}+\delta_{\rm  int}.
 \label{delgr}
\ee
In order to define the bias consistently on super-Hubble scales \citep{Challinor:2011bk,Bruni:2011ta,Jeong:2011as}, we use the comoving-synchronous matter overdensity 
\be
\dco=\delta -3aHV,
\ee
where $a=1/(1+z)$ is the scale factor. The local and integrated relativistic terms are
\begin{align}
\delta_{\rm loc}&=  \left(3-b_e\right)\frac{H}{(1+z)} V + An^i \partial_i V +(2\mb-2-A)\Phi + \frac{\dot\Phi}{H} \label{delgr1},\\
\delta_{\rm  int}&= 4  \frac{\left(1-\mb\right) }{\chi}\int_0^{\chi} {d\tilde \chi}\Phi  - 2A \int_0^{\chi} {d\tilde \chi} \frac{\dot\Phi}{(1+z)},
\label{delgr2}
\end{align}
where $b_e$ is the evolution bias, giving the source count evolution
\be
b_e=\frac{\partial\ln a^3N_g}{\partial\ln a},
\ee
and the factor $A$ is
\be
A=b_e -2\mb-1-\frac{\dot{H}}{{H}^2}+\frac{2\left(\mb-1\right)(1+z)}{\chi {H}}.
\ee

The local term $\delta_{\rm loc}$ has Doppler and Sachs-Wolfe type contributions. The integrated term $\delta_{\rm  int}$ contains time-delay and integrated Sachs-Wolfe contributions. These relativistic terms can become significant near and beyond the Hubble scale. (See \citealt{Yoo:2009au,Yoo:2010ni,Jeong:2011as,Bertacca:2012tp,Raccanelli:2013dza} for the local terms and \citealt{Bonvin:2011bg,Challinor:2011bk,Raccanelli:2013gja} for the integrated terms.) The growth of relativistic effects occurs on the same scales where the effect of PNG is growing through the galaxy bias of Eq.~\eqref{eq:bias-NG}. The relativistic effects are easily confused with the PNG contribution \citep{Bruni:2011ta,Jeong:2011as,Yoo:2012se,Raccanelli:2013dza}. In order to remove this theoretical systematic, it is necessary to include all the relativistic effects in an analysis of PNG in galaxy surveys \citep[see][for surveys in the radio continuum]{Maartens:2012rh}.

The computation of the observed number density contrast with all relativistic contributions can be done with \textsc{camb}\_sources \citep[][which we use here]{Challinor:2011bk} or \textsc{class}gal \citep{DiDio:2013bqa}. These codes compute the angular power spectra $C_\ell(z_i,z_j)$, which correspond to what is observed on the lightcone.

\section{SKA HI Galaxy Redshift Surveys}
The SKA project is set to be the largest radio telescope ever built, coordinating efforts from several countries and sharing facilities between South Africa and Australia. It is planned to be developed in two main phases. Phase 1 currently encompasses two mid-frequency facilities ($\sim 1$ GHz) operating within South Africa (SKA1-MID) and Australia (SKA1-SUR). A low frequency array (SKA1-LOW $\sim 100$ MHz) will be set in Australia. We refer to \citep{Dewdney:2009} for a description of the set-ups. In the second phase, the full SKA, the plan is to extend the array by about a factor of 10 in one of the sites, both in collecting area and primary beam (field of view), thus significantly increasing the survey power of the facility. 

One of the key science goals for the SKA will be the detection of a large amount of HI galaxies up to high redshifts using the 21 cm line of neutral hydrogen to measure very accurate redshifts. Cosmological applications will require detecting large numbers of galaxies to beat shot noise and over a large sky area to reduce cosmic variance. With the sensitivities for SKA1 and taking 10,000 hours of observing time, the optimal survey area will be around 5,000 deg$^2$ using band 2 from SKA1-MID ($> 950$ MHz) or SKA1-SUR ($> 650$ MHz). This will enable the detection of about $5\times 10^6$ galaxies up to $z\sim0.5$, with a flux cut five times above a flux sensitivity of $\sim70$ $\mu$Jy. The full SKA, on the other hand, should be capable of detecting about $9\times 10^8$ galaxies over a 30,000 deg$^2$ area, up to $z\sim2.0$, with a flux cut ten times above a flux sensitivity of $\sim5$ $\mu$Jy, making it the largest galaxy redshift survey ever.

\begin{figure}
\centering
\includegraphics[width=0.5\textwidth]{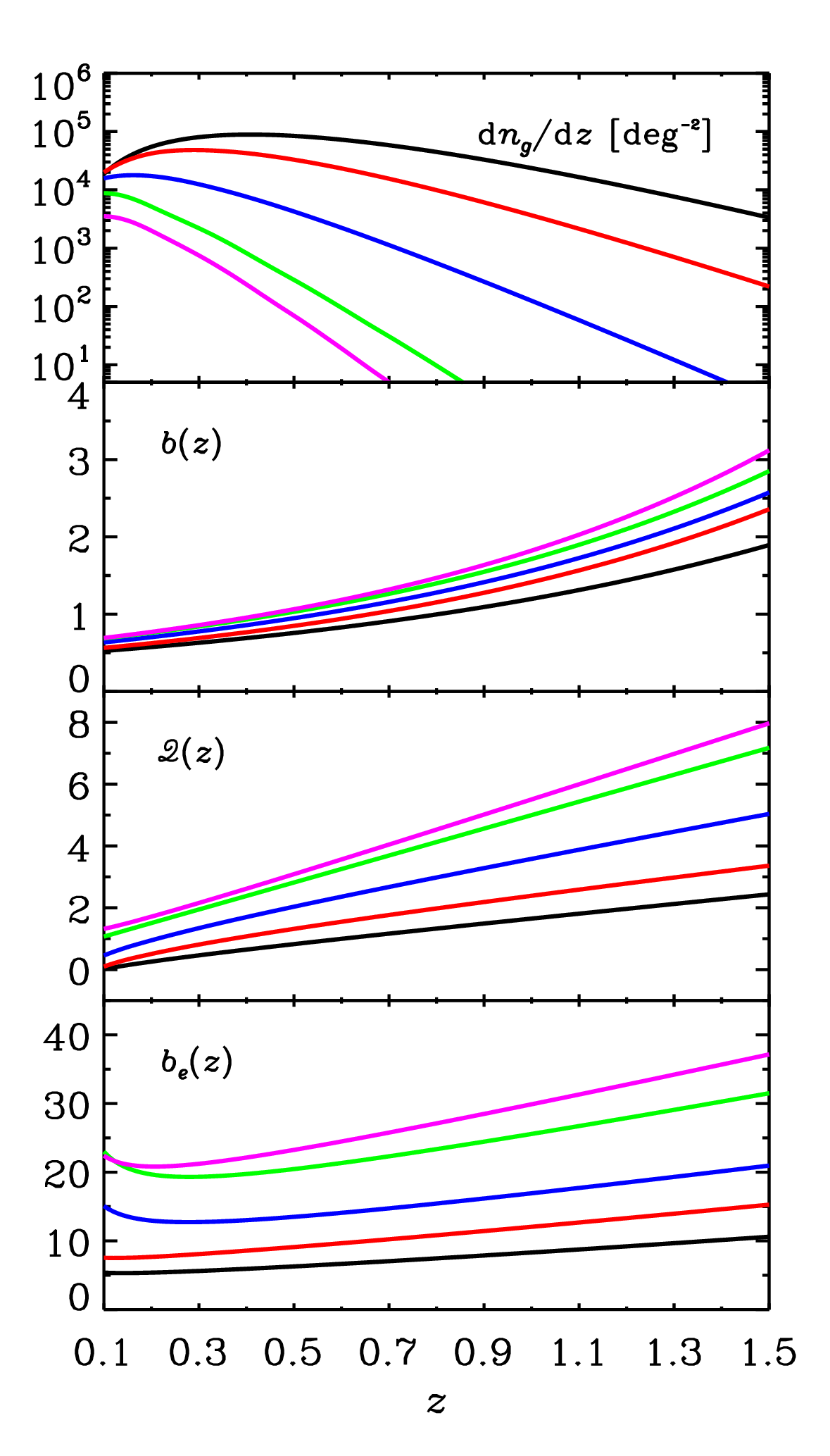}
\caption{Key functions derived from simulations for SKA HI galaxy redshift surveys, with sensitivities of $3$, $7.3$, $23$, $70$ and $100$ $\mu$Jy (black, red, blue, green and magenta curves, respectively). \textit{Top:} \textsc{Hi} galaxy number distribution per unit redshift per square degree, $\de n_g/\de z=a^3N_g(\chi^2/H)(\pi/180)^2$. \textit{Second:} HI galaxy bias, $b(z)$, for $\fnl=0$. \textit{Third:} Magnification bias, $\mb(z)$. \textit{Bottom:} Evolution bias, $b_e(z)$. [{\it Note:} in the published version (arXiv v4), there are errors in the $\mathcal Q$ and $b_e$ plots, which have been corrected here.] }\label{fig:mb}
\end{figure}

The noise calculations and parameters for these HI galaxy surveys can be found in \citet{Yahya:2014yva}. We used the SAX-sky simulation to calculate the HI galaxy number density and bias as a function of flux sensitivity.\footnote{\texttt{http://s-cubed.physics.ox.ac.uk/s3\_sax}} The results are summarised in Fig.~\ref{fig:mb}. Note that the flux numbers quoted in Fig.~\ref{fig:mb} refer to noise sensitivities---as explained above, the actual flux threshold is assumed to be five times this number for a SKA1 survey and ten times for the full SKA.

The numbers obtained below (presented in Sec.~\ref{ssec:results}) show that a galaxy redshift survey with SKA1 will not improve on {\it Planck} for PNG constraints. An HI intensity mapping survey has been proposed as a way to surpass this limitation with SKA1 and push the limits on cosmological constraints \citep[see][for details]{Bull:2014rha}. However, this type of survey will rely on sophisticated foreground cleaning techniques. The advantage of an HI galaxy redshift survey is that it will be much less contaminated by foregrounds or calibration artefacts, making the detection of the signal much cleaner. This is particularly important on very large scales where we are probing PNG. Our results show that as sensitivity is improved towards the full SKA, the number of galaxies detected will allow unmatched constraints on most cosmological parameters, including PNG.

\section{Forecasting Constraints on PNG}
The  observable that we employ in this analysis is the tomographic angular power spectrum of HI galaxy number density fluctuations \citep{Challinor:2011bk}, 
\begin{equation}
C^{ij}_{\ell}=4\pi\int\!\!\de\ln k\,\mathcal W^{{i}}_{\ell}(k)\mathcal W^{{j}}_{\ell}(k)\Delta^{2}_{\zeta}(k).\label{eq:C_l}
\end{equation}
Here the index $i$ refers to the $i$th tomographic redshift bin, which the redshift distribution of HI galaxies, $\de n_{g}(z)/\de z$, has been divided into, $\ell$ is the Legendre multipole of the spherical Bessel-Fourier decomposition, $k$ is the three-dimensional wavenumber, and $\Delta^{2}_{\zeta}(k)$ is the dimensionless power spectrum of primordial curvature perturbations. Eq.~\eqref{eq:C_l} describes the same observable as the angular power spectrum $C_\ell(z_i,z_j)$ mentioned in Sec.~\ref{ssec:gr}. However, we use from now on the notation of \eqref{eq:C_l} to emphasise that we bin the galaxy redshift distribution into $i,j=1,\ldots,N_z$ slices and then consider the angular power spectrum between the $i$th and $j$th bins, rather than picking up two infinitesimal redshift surfaces at $z_i$ and $z_j$, which is unfeasible for galaxy surveys. (Note that this is not the case for intensity mapping, see e.g.\ \citealt{Camera:2013kpa}.)

The window functions in \eqref{eq:C_l} are derived from binning  $\mathcal W_{\ell}(z,k)$, and contain the redshift distribution of the galaxy population and the transfer function for the galaxy number over-density.
In the standard Kaiser analysis, we have 
\begin{multline}
\mathcal W^{\rm std}_{\ell}(k)=\int\!\!\de\chi\,\frac{\de n_{g}(\chi)}{\de\chi}j_{\ell}(k\chi)\\\left\{[b[\chi(z)]+\Delta b[\chi(z),k]+f[\chi(z)]\mu_k^2\right\}T_\delta[\chi(z),k],\label{ws}
\end{multline}
where $j_{\ell}$ is a spherical Bessel function, $f=\de \ln D/\de \ln a$ is the growth function and $\mu_k=\mathbf k\cdot \mathbf n/k$. 

We want to perform for the first time a theoretically complete and self-consistent analysis of PNG forecasts, including all the relativistic effects described in Sec.~\ref{ssec:gr}, which are known to be important on horizon scales where the PNG signal is also growing. In the fully relativistic analysis, we need to modify Eq.~\eqref{ws} to include the additional terms in Eq.~\eqref{delgr}, beyond the Kaiser term. As mentioned before, to compute the full relativistic angular power spectrum of Eq.~\eqref{eq:C_l}, we use the \textsc{camb}\_sources package \citep{Challinor:2011bk}, which we modified in order to include the non-Gaussian bias correction of Eq.~\eqref{eq:bias-NG}.

To forecast the SKA potential for constraining PNG, we perform a Fisher matrix analysis \citep{Fisher:1935,Tegmark:1996bz} focussed on the PNG parameter $\fnl$. If we assume that the model likelihood surface in parameter space can be well approximated by a multivariate Gaussian (at least in a neighbourhood of its peak), the Fisher matrix $\mathbfss F$ is then a good approximation for the inverse of the parameter covariance. This means that for a set of parameters $\{\vartheta_\alpha\}$ we can write
\be
\mathbfss F_{\alpha\beta}=\fsky\sum_{\ell=\lm}^\lM\frac{2\ell+1}{2}\frac{\partial C^{ij}_{\ell}}{\partial \vartheta_\alpha} \Sigma^{jp}_{\ell}\frac{\partial C^{pq}_{\ell}}{\partial \vartheta_\beta} \Sigma^{qi}_{\ell},\label{eq:fisher}
\ee
where we assume summation over repeated indices, and $\Sigma^{ij}_{\ell}$ is the inverse of $C^{ij}_{\ell}+\mathcal N^{ij}_{\ell}$, due to the fact that measurements of $C^{ij}_{\ell}$ are affected by Poisson noise,
\be
\mathcal N^{ij}_{\ell}=\frac{\delta^{ij}}{\bar N_g^i}.\label{eq:noise}
\ee
Here $\bar N_g^i$ is the total number of galaxies per steradian in the $i$th redshift bin, and $\fsky$ is the fraction of the sky covered by the survey, viz.\ $\fsky\equiv {\rm Area}/(4\pi)$, corresponding to 0.73 for 30,000 deg$^2$ and 0.12 for 5,000 deg$^2$.

If all the model parameters were uncorrelated with each other, the forecast error on parameter $\vartheta_\alpha$ would simply be (no equal index summation)
\be
\Delta(\vartheta_\alpha)=\frac{1}{\sqrt{\mathbfss F_{\alpha\alpha}}}.\label{eq:conditional}
\ee
In Bayesian statistics this is known as `conditional error'. On the other hand, to take parameter degeneracies properly into account, one should always consider `marginal errors', i.e.
\be
\sigma(\vartheta_\alpha)=\sqrt{\left(\mathbfss F_{\alpha\alpha}\right)^{-1}}.\label{eq:marginal}
\ee
In the following we shall use Eq.~\eqref{eq:marginal} and only occasionally refer to the conditional errors of Eq.~\eqref{eq:conditional}. 

For the choice of parameters, one should in principle consider the full \lcdm\ parameter set. One possibility to account for our current knowledge on \lcdm\ parameters is to add afterwards to the full Fisher matrix a matrix of priors coming from external experiments. Another way, widely adopted in the literature \citep[e.g.][]{Sartoris:2010cr,Fedeli:2010ud,Fedeli:2012dg,Ade:2013ydc,Raccanelli:2014kga,Raccanelli:2014awa}, is to vary freely only the parameters one is interested in and fix all the others to their fiducial values, as fitted for instance by \textit{Planck}. On the one hand, this possibly leads to an overestimation of a survey constraining power, for by doing so one may as well disregard parameter degeneracies. However, it is known that the inclusion of PNG does not significantly broaden other parameter confidence intervals \citep{Giannantonio:2011ya}.

\section{Forecast Results}\label{ssec:results}
In order to forecast constraints for  $\fnl$,  we proceed as follows. We limit ourselves to a multipole range $2\leq \ell \leq700$ \citep[as also done by][]{Ferramacho:2014pua}, where non-linear corrections to galaxy clustering may be safely neglected. On these angular scales, the cosmological parameter that can be most degenerate with $\fnl$ is the normalisation of the matter power spectrum, $\sigma_8$. Hence, we adopt $\vartheta_\alpha=\{\fnl,\sigma_8\}$ and leave the other parameters fixed. (We discuss the impact of this assumption in Sec.~\ref{ssec:degeneracies}.) 

\subsection{Constraints for $z\leq1.5$}
We limit ourselves for now to $z\leq1.5$. This corresponds to the redshift range that most galaxy surveys will be able to probe. We divide the HI galaxy redshift distribution into 20 equally spaced bins in the range $0<z\leq1.5$. Note that we fully account for auto- and cross-correlations amongst all redshift bins.

Table~\ref{tab:constraints} summarises the constraints that we forecast for various SKA experimental set-ups, which can be easily related to SKA1 or full SKA configurations. It is apparent from the third and fourth columns that our assumption regarding the parameter set appears to be well founded. Indeed, we see that for the highest sensitivities (lowest flux cuts), the conditional and marginal errors on $\fnl$ coincide. From Eq.~\eqref{fpg} it follows that SKA HI galaxy redshift surveys will be able to achieve more than twice the accuracy of \textit{Planck} in a measurement of $\fnl$, using the redshift range  $0<z\leq1.5$, i.e.\
\be
\sigma(\fnl)=3.12~~~(0< z\leq1.5).
\ee
\begin{table}
\caption{Summary of forecast constraints on $\fnl$ for $z\leq1.5$. The third and fourth columns refer respectively to conditional errors and errors marginalised over $\sigma_8$.}\label{tab:constraints}
\centering
\begin{tabular}{cccc}
Sensitivity [$\mu$Jy] & $f_{\rm sky}$ & $\Delta(\fnl)$ & $\sigma(\fnl)$ \\
\hline
3 & 0.73 & 3.12 & 3.12 \\
7.3 & 0.73 & 3.23 & 3.23 \\
23 & 0.73 & 7.80 & 8.30 \\
70 & 0.12 & 18.5 & 24.6 \\
100 & 0.12 & 53.3 & 79.3
\end{tabular}
\end{table}

\begin{figure*}
\centering
\includegraphics[width=0.9\textwidth]{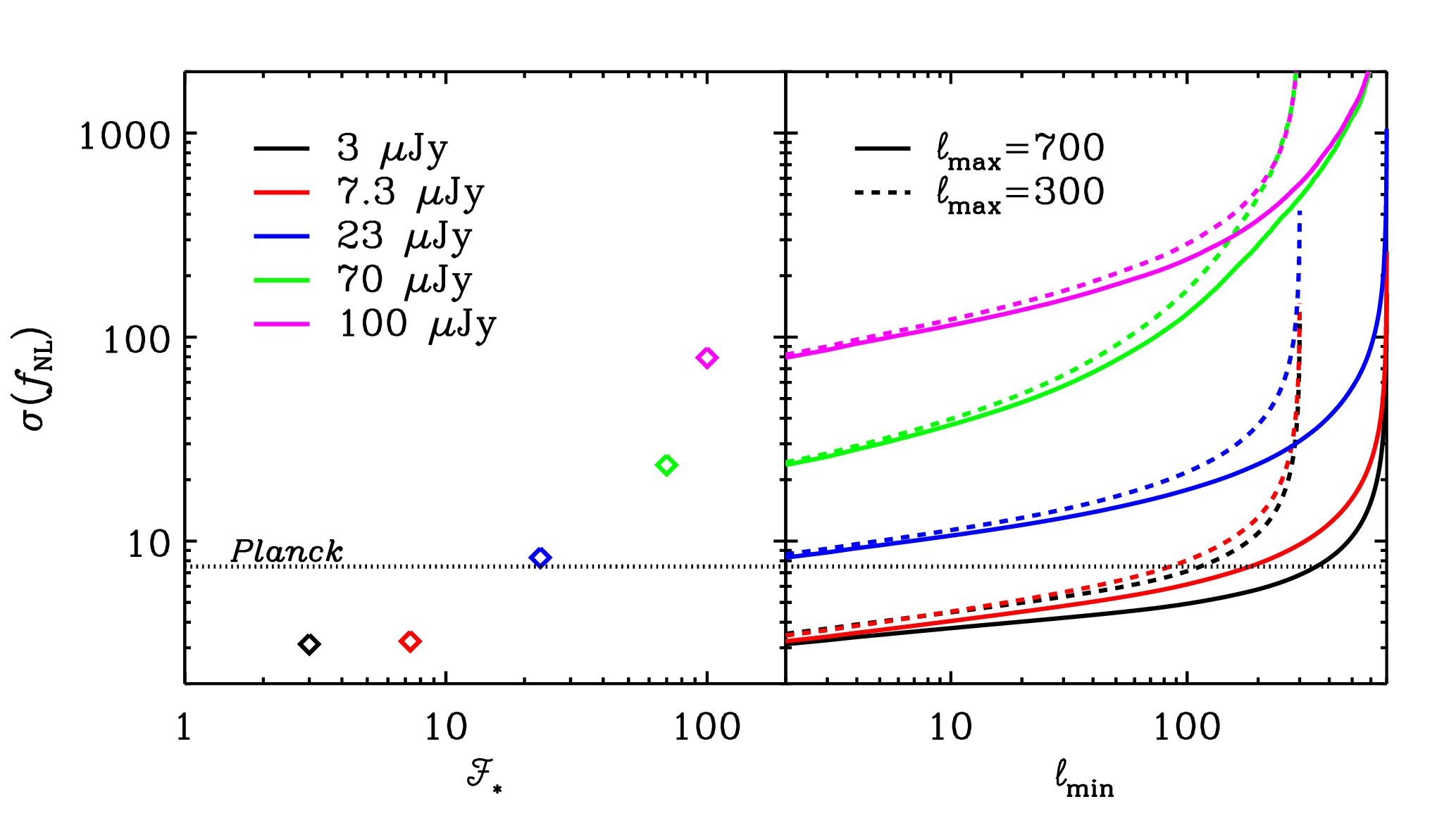}
\caption{Forecast 68$\%$ errors on $\fnl$ marginalised over $\sigma_8$ for $0<z\leq 1.5$. \textit{Left:} As a function of flux cut for $\lm=2$ and $\lM=700$. The dotted horizontal line shows the current \textit{Planck} constraint (in the LSS convention). \textit{Right:} As a function of $\lm$ and for various flux cuts and two values of $\lM$.}\label{fig:sfNL}
\end{figure*}

The left panel of Fig.~\ref{fig:sfNL} shows the fourth column of Table~\ref{tab:constraints}. In the right panel, we illustrate the dependence of $\sigma(\fnl)$ on the minimum angular multipole considered in the analysis, $\lm$. Again, we present this for all the flux cuts discussed above, as well as for a more conservative value of the maximum angular multipole, namely $\lM=300$. The constraints weaken as expected, when we increase $\lm$, since it is on the largest angular scales that PNG has the strongest impact. However, the choice of $\lM$ does not influence much the final outcome. Indeed, a more conservative value like $\lM=300$ yields almost the same results.

\subsection{Pushing to higher redshifts}\label{ssec:volume}
It has been shown (see for instance \citealt{Camera:2013kpa} or \citealt{Ferramacho:2014pua}) that to probe PNG effectively, one needs to access as big a `cube' of Universe as possible, in order to include the largest modes and lessen cosmic variance. (Moreover, if possible we can use a multi-tracer analysis \citep{Seljak:2008xr} with two or more proxies of the underlying dark matter distribution.) 

Large volumes, with redshift information, can in principle be delivered by a SKA1 HI intensity mapping survey \citep[see e.g.][]{Bull:2014rha}, but several technical issues still need to be addressed. SKA galaxy redshift surveys will achieve exquisite accuracy in the measurement of HI emitting galaxies, cataloguing almost one billion such sources with precise redshift estimates. As can be seen in Fig.~\ref{fig:mb}, the redshift distribution of SKA HI galaxies has a non-negligible high-redshift tail, when high sensitivities such as 3 $\mu$Jy are achieved. This is an advantage of HI surveys compared to other galaxy redshift surveys, given the abundance of HI at high $z$.

By exploiting this, we can explore the capabilities of SKA HI galaxy redshift surveys in accessing the largest scales at higher redshifts. The outcome of such an analysis is not straightforward to predict. On one hand, the higher the redshift the larger the bias (since we can probe smaller $k$'s in the linear regime), and thus the stronger the PNG signal.  On the other hand, the decrease in the number of sources as we go to high $z$ reduces the constraining power because of the Poisson noise term in Eq.~\eqref{eq:noise}.

We consider the case of a 3 $\mu$Jy flux sensitivity and perform the same Fisher analysis described before but this time considering the range $1.5\leq z\leq3$, again subdivided into 20 equally spaced bins. Although the noise is indeed larger in this case compared to the case of the previous section, we nevertheless find a tighter constraint
\be
  \sigma(\fnl)=1.88~~~(1.5\leq z\leq3) .
\ee   
This is a very promising result, and represents the most stringent constraint forecast for a future galaxy redshift survey exploiting one single tracer. 

In fact, we can go further, and combine the two redshift chunks. The proper way to do so would be to bin $C^{ij}_{\ell}$ in the whole $0<z\leq3$ interval. However, the calculation of a large number of redshift bins and all their cross-correlations is computationally cumbersome. To overcome this problem, we follow \citet{Camera:2013kpa} and construct a 40$\times$40 block-diagonal tomographic $C^{ij}_{\ell}$ matrix, then correct for the disregarded cross-correlations by overlapping a further 20-bin square matrix for the intermediate range $0.75\leq z\leq2.25$. This allows us to put stronger bounds on $\fnl$:
\be
\sigma(\fnl)=1.54~~~(0<z\leq3).
\ee
This constraint is five times tighter than the current \textit{Planck} constraint. Furthermore, it is tight enough for testing slow-roll single-field inflation.

\begin{figure}
\centering
\includegraphics[width=0.5\textwidth]{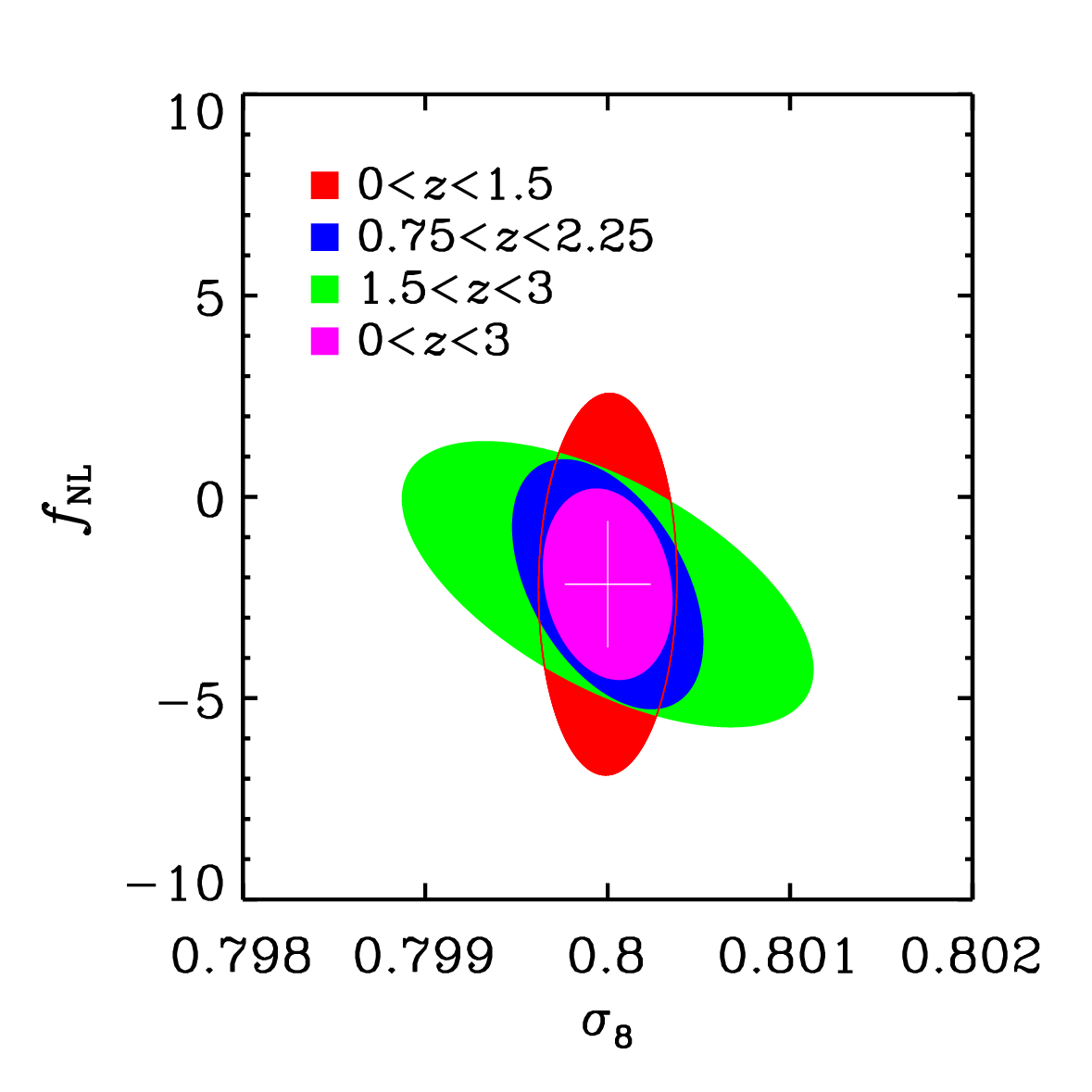}
\caption{Marginal 1$\sigma$ error contours in the $\sigma_8$--$\fnl$ plane at  3$\,\mu$Jy sensitivity for the three redshift chunks and their combination.}\label{fig:ellipses}
\end{figure}

Figure~\ref{fig:ellipses} illustrates this major result by showing the marginal 1$\sigma$ error contours in the $\sigma_8$--$\fnl$ plane for the three redshift chunks separately (red, blue and green ellipses) and for their combination (magenta contour). We can explain the different orientations and sizes of the ellipses as follows.

Firstly, as we probe higher redshifts, constraints on $\sigma_8$ broaden. This is because the matter power spectrum decreases in amplitude as $z$ increases, and so the sensitivity of $C^{ij}_{\ell}$ to $\sigma_8$ too decreases with redshift. 

Secondly, let us focus on the $\fnl$-axis, where there are more subtle effects. It appears as though the most stringent constraint is obtained in the middle redshift interval. Indeed, we get $\sigma(\fnl)=3.12$, $1.82$ and $1.88$ for the first, second and third chunk, respectively. At first glance, this seems to contradict the fact that PNG deviations from the Gaussian prediction monotonically increase with redshift. However, we must not forget that Poisson noise also grows with $z$. As a result, the middle redshift interval is the one where the balance between PNG effects and noise is more optimal. A further check can be made by comparing the forecast constraints on $\fnl$ in the idealistic case of zero noise. We find that $\sigma(\fnl)=2.82$, $1.29$ and $0.75$ for the first, second and third chunk, respectively---in agreement with our qualitative expectations. All these major results are summarised in Table~\ref{tab:high-z_constraints}, where we denote by $\widehat\sigma(\fnl)$ the marginal errors in the 
noiseless case.
\begin{table}
\caption{Forecast constraints on $\fnl$ marginalised over $\sigma_8$ for 3$\,\mu$Jy sensitivity and the redshift ranges discussed in Sec.~\ref{ssec:volume}, with and without noise (second and third column, respectively).}\label{tab:high-z_constraints}
\centering
\begin{tabular}{rclcc}
\multicolumn{3}{c}{Redshift range} & $\sigma(\fnl)$  & $\widehat\sigma(\fnl)$ \\
\hline
$0<\hspace{-0.1in}$ & $z$ & $\hspace{-0.1in}\leq1.5$ & 3.12 & 2.82 \\
$0.75\leq\hspace{-0.1in}$ & $ z$ & $\hspace{-0.1in}\leq2.25$ & 1.82 & 1.29 \\
$1.5\leq\hspace{-0.1in}$ & $ z$ & $\hspace{-0.1in}\leq3$ & 1.88 & 0.75 \\
$0<\hspace{-0.1in}$ & $z$ & $\hspace{-0.1in}\leq3$ & 1.54 & 0.71
\end{tabular}
\end{table}

\subsection{PNG parameter degeneracies}\label{ssec:degeneracies}
Our analysis assumed that all parameters were fixed except for $\fnl$ and $\sigma_8$. This could lead to some concerns about how this assumption might affect the constraints on $\fnl$. However, there are two important points that should be taken into account and which give us some security that the analysis is solid. Firstly, most cosmological parameters do not really have an effect on the large scales (small $\ell$) that we are considering. Moreover, small effects on these scales should be easy to disentangle from the $k^{-2}$ signature that one expects for PNG. One such example is $\sigma_8$, which only shifts the power spectrum by an overall amplitude. Secondly, most of these same cosmological parameters should be very well measured at other scales and in particular using CMB data, which is independent of the $\fnl$ effects. Summarising, not only we do expect most parameters to be very well constrained already, but their effects on large scales should be negligible or easily distinguishable from the $k^{-2}
$ signature we are looking for.

Two issues however need a more detailed analysis: the effect of the galaxy bias $b$ and the effect of the extra parameters such as $\mathcal Q$ and $b_e$ in the relativistic corrections to $C^{ij}_\ell$. Changes in these parameters could in principle mimic $\fnl$ effects, since they also can affect power on large scales [the bias does this through the $(b-1)\fnl$ term in Eq.~\eqref{eq:bias-NG}].

However, as above, $b(z)$ should be reasonably well constrained by measurements at smaller (but still linear) scales, using a combination of RSDs and the overall amplitude of the power spectrum (where we can safely neglect the possible $k^{-2}$ dependence). We do need to assume that the bias is scale-independent so that we can use the same value on large scales, but this should be reasonable in the linear regime. The information from the CMB will further lift any degeneracies. Moreover, even if small uncertainties in the bias still exist, the fact that there should be a reasonable monotonic evolution with redshift while $\fnl$ is constant can be used to further disentangle any remaining degeneracies.

A very pessimistic approach would be to consider the bias as a free parameter in each redshift bin. However, we do not expect this to degrade considerably our constraints since the measurement of $\fnl$ relies on the ability to pick up the $k^{-2}$ dependence; this should thus not be affected by the uncertainty in the bias as long as it is scale independent on these scales. Note also that we have marginalised over $\sigma_8$ which is equivalent to marginalising over an overall constant bias.

\begin{figure}
\centering
\includegraphics[width=0.5\textwidth]{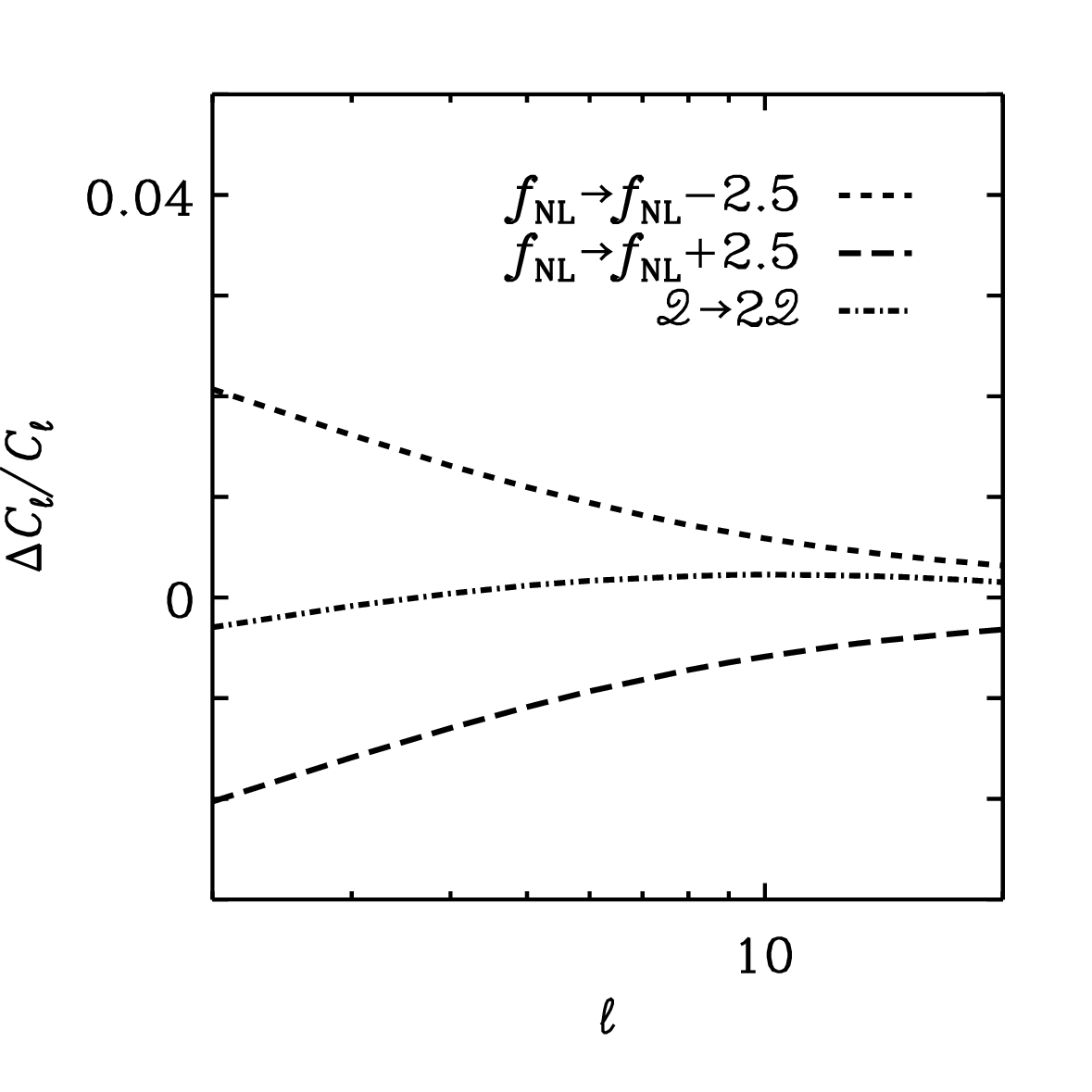}
\caption{Ratio of the angular power spectrum to the fiducial one for the third redshift slice of the 3 $\mu$Jy case when $\fnl$ and $\mb$ are changed.}\label{fig:ratios}
\end{figure}

Regarding the relativistic parameters, i.e.\ the magnification bias $\mb$ and the evolution bias $b_e$, the situation is similar. They both depend on the total galaxy number counts which should be very well measured at small scales using deep galaxy surveys. Obviously we need to be careful with possible cosmic variance effects. Even if there remain small uncertainties in $\mb$ or $b_e$, they should not impact on the extraction of $\fnl$. We illustrate this in the case of $\mb$ in Fig.~\ref{fig:ratios}. This shows that the effect of $\mb\to 2\mb$ is much smaller than the effect of changing $\fnl$ within the 1$\sigma$ limits of our constraints.

\section{Conclusions}
In this paper, we made a detailed analysis of the expected PNG constraints from a HI galaxy survey, by using its effect on the large scale power spectrum and fully taking into account the relativistic corrections expected at these scales. The relativistic corrections are small compared to the effects of astrophysical modelling, but this is not the relevant comparison: the point is that the relativistic effects are comparable to the PNG signal. By omitting these effects and using the Newtonian-Kaiser approximation, one introduces an important theoretical systematic. We avoid this problem by using the correct relativistic analysis.

There are two aspects to the relativistic effects.
\begin{itemize}
\item The first is an important non-linear correction to the Poisson constraint which induces a correction $\fnl\to \fnl -2.17$ when using LSS observations. This changes the fiducial value of $\fnl$ from 0 to $-2.17$, and, more importantly, it shows that the simplest inflation models would give $\fnl \simeq -2.2$ in LSS and $\fnl \simeq 0$ in the CMB. We have shown that HI galaxy redshift surveys with the full SKA are able, in combination with the CMB, to detect the PNG signal from the simplest inflation models.\\
\item The second is the linear corrections to the observed galaxy number counts and power spectrum. The effect on $\sigma(\fnl)$ is typically small, but there will be a bias in the best-fit estimates of $\fnl$ from observations, given that the relativistic effects are comparable to the PNG signal.
\end{itemize}

Measuring the PNG signature on the clustering bias of dark matter tracers has the advantage that it is free from non-linearities that can affect the $\fnl$ measurements on smaller scales using the bispectrum estimator. Furthermore, it will not be subject to foreground residuals or calibration artefacts that might contaminate the higher order correlation functions commonly used to estimate the amount of PNG.

In addition to HI galaxy (threshold) surveys, radio telescopes can also conduct HI intensity mapping surveys, which do not detect individual galaxies but the total HI emission, and are therefore ideally suited to map the large-scale distribution relevant for PNG. The intensity mapping surveys cover the same sky area but to greater depth (or cover the same depth in less time). This is shown to be particularly advantageous for a HI intensity mapping survey with SKA1 \citep{Camera:2013kpa}. On the other hand, galaxy surveys should give a pristine signal on these scales while foreground cleaning issues and correlated noise effects are still not completely clarified for HI intensity mapping surveys.

Our results show that a survey with Phase 1 of the SKA should reach 1$\sigma$ constraints on $\fnl$ of about 20. This is mostly due to the quick drop with redshift of the expected galaxy number density at this flux levels. On the other hand, as we move towards the full SKA, it will be possible to make a detection below $\fnl\sim2$, thanks to the large number of HI galaxies that will be detected up to high $z$. This will be 5 times better than the best current constraints (from {\it Planck}) and will start to probe into `standard' inflationary model physics. Other galaxy surveys such as the European Space Agency \textit{Euclid} satellite\footnote{\texttt{http://www.euclid-ec.org/}} \citep{Laureijs:2011gra,Amendola:2012ys} will also target this type of measurement but only the full SKA should be able to probe to such high redshifts ($z\sim 3$) over large areas of the sky, thanks to the abundance of HI galaxies as we move to higher redshifts, compared to other lines.

Finally we would like to emphasise that the numbers we used for the bias and galaxy redshift distribution for different flux cuts were taken directly from simulations. The other parameters required for the relativistic corrections were in turn directly derived from these parameters, which allowed us to provide a fully consistent analysis of the fluctuations on ultra-large scales taking into account the general relativistic corrections. The obtained constraints should be reasonably insensitive to changes in other parameters due to the specific $k^{-2}$ signature from PNG as well as the fact that most parameters should be already measured from surveys at smaller scales and the CMB itself.

{\it NOTE ADDED.} ArXiv v4 of this paper is the version published in MNRAS. The corrections in this version are published as an Erratum (see Journal reference for details).

\subsection*{Acknowledgments}
We thank Daniele Bertacca, Phil Bull, Benjamin Joachimi, Kazuya Koyama, Sabino Matarrese, Alvise Raccanelli and Licia Verde for useful clarifications and discussions, and Cristiano Porciani for alerting us to an error in arXiv v4. SC acknowledges support from FCT-Portugal under Post-Doctoral Grant No. SFRH/BPD/80274/2011. MS and RM are supported by the South African Square Kilometre Array Project and the South African National Research Foundation. RM is also supported by the UK Science \& Technology Facilities Council, Grant No. ST/K0090X/1.

\bibliographystyle{mn2e}
\bibliography{../../../Bibliography}

\newpage
\appendix
\section{Fitting Formulas for Magnification Bias}
\citet{Yahya:2014yva} provides fitting formulas for $\de n_g(z)/\de z$ per square degree as a function of redshift for various flux sensitivities. In this appendix, we give similar fittings for $\ln N_g(z,\mathcal F>\mathcal F_{\ast})$ as a function of flux cut at a given redshift, useful for evaluating $\mathcal Q(z,\mathcal F_{\ast})$ [see Eq.~\eqref{mbq}].

\begin{figure}
\centering
\includegraphics[width=0.5\textwidth]{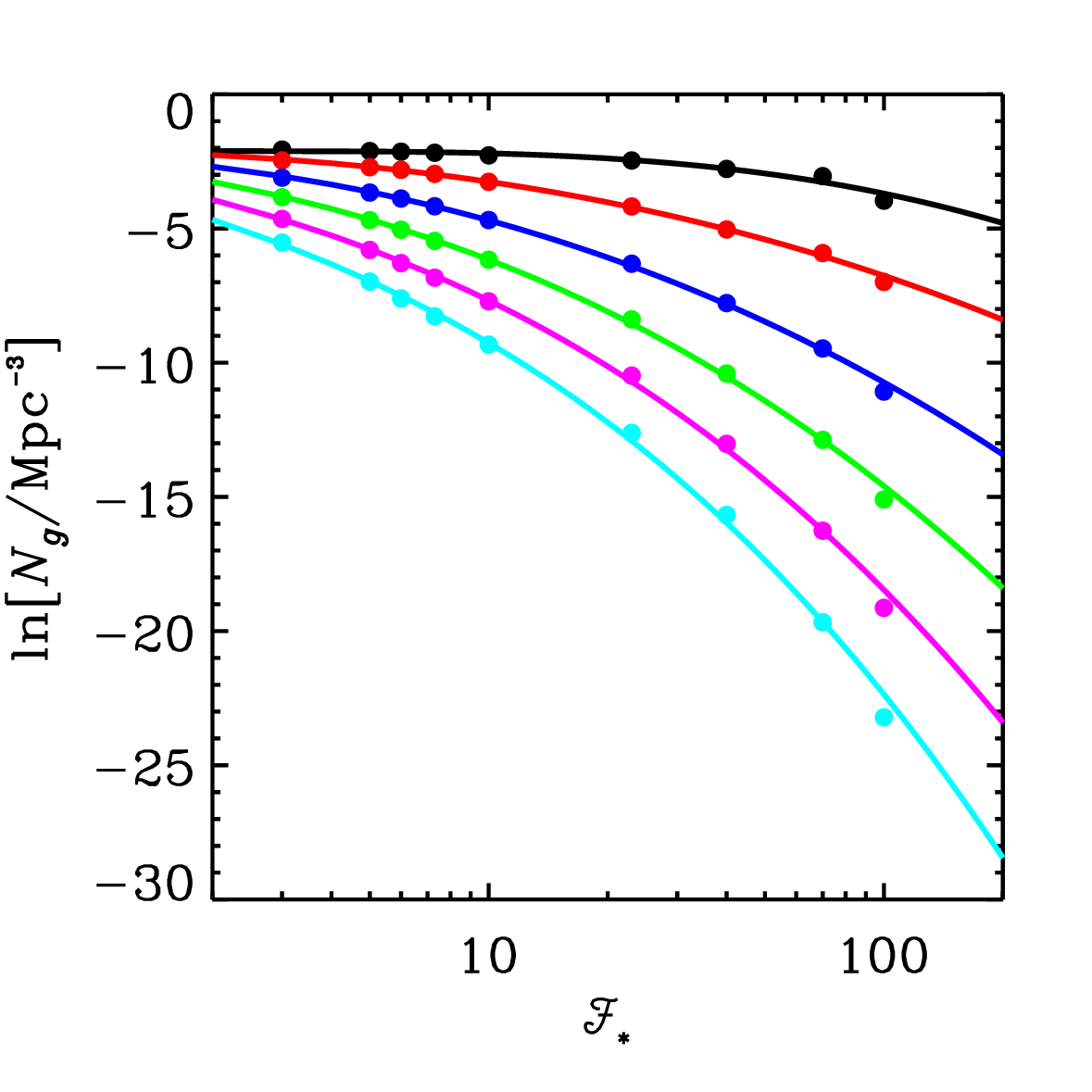}
\caption{Background galaxy number density with flux larger than the flux threshold for $z=0.1$, $0.3$, $0.6$, $0.9$, $1.2$ and $1.5$ from top (right) to bottom. Solid lines are the outcome of Eq.~\eqref{eq:fit}, bullets come from the tables in \citet{Yahya:2014yva}. [{\it Note:} This corrects errors in the published version (arXiv v4).]}\label{fig:fit}
\end{figure}

We propose the formula\footnote{This changes the equation in the published version (arXiv v4).}
\begin{equation}
\ln N_g(\mathcal F>\mathcal F_{\ast}\!)=-a_{0}-a_{1}\ln\mathcal F_{\ast}-a_{2}\left(\ln\mathcal F_{\ast}\!\right)^{2}-a_{3}\left(\ln\mathcal F_{\ast}\!\right)^{3}\!,\label{eq:fit}
\end{equation}
with values of the fitted coefficients $a_1$, $a_2$ and $a_3$ given in Table~\ref{tab:fit} for various redshifts in the range $0<z\leq1.5$. (Intermediate redshift values can be obtained by interpolating.) As is clear from Fig.~\ref{fig:fit}, our fit is in very good agreement with the tabulated $\de n_{g}(z,\mathcal F_{\ast})/\de z$, with errors on average below a few percent. Note that Eq.~\eqref{eq:fit} contains no coefficient for the linear term in $\ln\mathcal F_{\ast}$. We found that the agreement was as good with a coefficient of $-1$ as with a free coefficient. Moreover, the correspondent $\mathcal Q$ is in good qualitative agreement with the phenomenological fitting formul\ae\ provided in the literature by \citet{Xinjuan:2011dw}.
\begin{table}
\caption{Coefficients of the fitting formula of Eq.~\eqref{eq:fit}. [{\it Note:} This corrects errors in the published version (arXiv v4).]} \label{tab:fit}
\centering
\begin{tabular}{lllll}
\multicolumn{1}{c}{$z$} & \multicolumn{1}{c}{$a_{0}$} & \multicolumn{1}{c}{$a_{1}$} & \multicolumn{1}{c}{$a_{2}$} & \multicolumn{1}{c}{$a_{3}$}\\
\hline
 0.1 & 2.0641 & 0.1280 & $ -0.1090$ & 0.0344 \\
 0.3 & 2.1181 & 0.1049 & 0.1428 & 0.0116 \\
 0.5 & 2.2197 & 0.2906 & 0.2238 & 0.0116 \\
 0.7 & 2.3676 & 0.5381 & 0.2541 & 0.0183 \\
 0.9 & 2.5579 & 0.8159 & 0.2596 & 0.0284 \\
 1.1 & 2.7867 & 1.1117 & 0.2505 & 0.0405 \\
 1.3 & 3.0504 & 1.4195 & 0.2315 & 0.0539 \\
 1.5 & 3.3455 & 1.7357 & 0.2056 & 0.0682
 \end{tabular}
\end{table}

\bsp

\label{lastpage}

\end{document}